\documentclass[aps,pre,preprint,showpacs,superscriptaddress]{revtex4-1}
\usepackage{ifpdf}
\usepackage{pdfsync}
\usepackage{hyperref}
\usepackage{graphicx,subfigure}
\usepackage{xcolor}
\usepackage{amsmath,amsfonts,amssymb}
\usepackage{braket}

\graphicspath{{figs/}}
\begin{document}

\title{Point-like Rashba interactions as singular self-adjoint extensions of the Schr\"{o}dinger operator in one dimension}
\author{V.L. Kulinskii}
\email{kulinskij@onu.edu.ua}
\affiliation{Department for Theoretical
Physics, Odessa National University, Dvoryanskaya 2,
65026 Odessa, Ukraine}
\author{D.~Yu. Panchenko}
\email{dpanchenko@onu.edu.ua}
\affiliation{Department for Theoretical
Physics, Odessa National University, Dvoryanskaya 2,
65026 Odessa, Ukraine}
\affiliation{Department of Fundamental Sciences, Odessa Military Academy, 10~Fontanska Road, Odessa 65009, Ukraine}
\begin{abstract}
We consider singular self-adjoint extensions for the Schr\"{o}dinger operator of spin-$1/2$ particle in one dimension. The corresponding boundary conditions at a singular point are obtained.
There are boundary conditions with the spin-flip mechanism, i.e. for these point-like interactions the spin operator does not commute with the Hamiltonian. One of these extensions is the analog of zero-range $\delta$-potential. The other one is the analog of so called $\delta^{(1)}$-interaction. We show that in physical terms such contact interactions can be identified as the point-like analogues of Rashba Hamiltonian (spin-momentum coupling) due to material heterogeneity of different types.
The dependence of the transmissivity of some simple devices on the strength of the Rashba coupling parameter is discussed. Additionally, we show how these boundary conditions can be obtained in the non-relativistic limit of Dirac Hamiltonian.
\end{abstract}
\maketitle
\section{Introduction}
Point-like interactions can be described as the singular extensions of the Hamiltonian and are very useful quantum mechanical models because of analytically tractability  \citep{book_demkovostrvsk_en,book_bazeldovichperelomov_en,math_deltalbergestexactsolv,funcan_adamyan_krein_jmathsci1988,funcan_alberverio_singperturb}. They are equivalent to some boundary conditions at the singular points and represent the limiting cases of field inhomogeneities. Therefore it is important to understand the relation between parameters of these BC and the specific physical characteristics of inhomogeneities. In  modern nanoengineering the spin control is of great interest \citep{qm_datta_advphys1990,qm_spintronic_ssc2001}. Besides the external magnetic field another interaction which could be used for such controlling is the spin-momentum coupling \citep{qm_rashbaorigin_physolid1960,qm_rashbabychkov_zhetp1984}.
The inclusion of magnetic field and other interactions which influence spin dynamics is a natural route for searching spin-dependent singular interactions. The interactions which influence spin polarization would give new examples of contact interactions with applications in condensed matter physics and QFT \citep{qm_deltasingnieto_physconferser2017}.
\section{Contact interactions for spin 1/2 case}\label{sec_spinx2x3}
In non relativistic limit spin $s=1/2$ particle is described by the Pauli Hamiltonian \cite{book_ll4_en}:
\begin{equation}\label{eq_ham_pauli}
\hat{H} = \frac{\left(\hat{p} - \frac{q}{c}\,\mathbf{A}\right)^2}{2\,m} + q\,\varphi - \frac{q\,\hbar}{2\,m\,c}\,\hat{\boldsymbol{\sigma}}\cdot\vec{\mathcal{H}}
\end{equation}
with $\boldsymbol{\sigma}$ representing the vector of Pauli matrices and $\vec{\mathcal{H}}$ is the external magnetic field with $\mathbf{A}$ is its vector potential and $\varphi$ is the scalar potential.
This Hamiltonian acts in space of 2-component wave functions:
where:
\begin{eqnarray}\label{psi_spin}
  \Psi = \left(
           \begin{array}{c}
             \psi_{\uparrow} \\
              \psi_{\downarrow} \\
           \end{array}
         \right)\,,
\end{eqnarray}
and $\psi_{\uparrow}, \psi_{\downarrow}$ are the wave functions of corresponding spin ``up-`` and ``down-`` states $\ket{\uparrow},\ket{\downarrow}$.
The probability current for \eqref{eq_ham_pauli} reads:
\begin{eqnarray}\label{j_spinpauli}
{\rm \mathbf{J}}_{w} = {\frac{\hbar} {m}}{\rm
Im}\left( {\Psi^{\dagger} \nabla \Psi} \right) - \frac{q}{mc}{\rm
\mathbf{A}}\Psi^{\dagger} \Psi + {\frac{\hbar}{2m}}{\rm rot}\left({\Psi^{\dagger} \boldsymbol{\sigma} \,\,\Psi}  \right)\,,
\end{eqnarray}
with the last term describing the magnetization current.

Bearing in mind the application to the 1-dimensional layered systems with spatial heterogeneity we use the conservation of current \eqref{j_spinpauli} to derive proper boundary conditions (BCs) for free particle with spin $s= 1/2$ modeling point-like interactions. We use the results of \cite{funcan_deltadistr_kurasov_jmathan1996} where all possible self-adjoint BCs were related with the following Hamiltonian:
\begin{equation}\label{kurasov_d2}
    L_X=-D_{x}^2\left(\, 1 + X_4\,\delta \,\right) + i\,D_{x}\left(\, 2\,X_3\,\delta - i\,X_4\,\delta^{(1)}\right) + X_1\,\delta+(X_2 - i\,X_3)\,\delta^{(1)}\,.
\end{equation}
Here symbol $D_x$ stands for the derivative in the sense of distributions on the space of functions continuous except at the point of singularity where they have bounded values along with derivatives \cite{funcan_deltadistr_kurasov_jmathan1996,funcan_deltafinitrank_procmath1998}:
\begin{equation}\label{kurasov_delta}
    \delta(\varphi) = \frac{\varphi(+0)+\varphi(-0)}{2}\,,\quad
    \delta^{(1)}(\varphi) = -\frac{\varphi'(+0)+\varphi'(-0)}{2}
\end{equation}
The parameters $X_i\in \mathbb{R}$ determine the values of the discontinuities of the wave function and its first derivative.
The boundary conditions (b.c.) corresponding to each contribution in Eq.~\eqref{kurasov_delta} can be represented in matrix form:
\begin{equation}\label{bc_matrixform}
    \left(
    \begin{array}{c}
        \psi(0+0) \\
        \psi'(0+0)
    \end{array}
    \right)
= M_{X_i}\,
    \left(
    \begin{array}{c}
        \psi(0-0) \\
        \psi'(0-0)
    \end{array}
    \right)
\end{equation}
and conserve the current (we put $\hbar=1,\, c=1$ and $m=1/2$):
\begin{equation}\label{j_free}
j = 2\,{\rm Im} \left(\psi^{*}\,\psi'\right)
\end{equation}
of the Hamiltonianfootnote{we put $\hbar=1$ and $m=1/2$}:
\begin{equation}\label{ham_free}
\hat{H} = - \frac{d^2}{d\,x^2}
\end{equation}
of a spinless particle. Physical classification of all these b.c. on the basis of  gauge symmetry breaking was proposed in \cite{qm_zrpsymme_physb2015}. They can be divided into tow subsets. The first one is formed by the matrices:
\begin{equation}\label{eq_bcxqx4}
M_{X_1}=\left(
        \begin{array}{cc}
          1 & 0 \\
          X_1 & 1 \\
        \end{array}
      \right)\,, \qquad
      M_{X_4}=\left(
        \begin{array}{cc}
          1 & - X_4 \\
          0 & 1 \\
        \end{array}
      \right)\,, \end{equation}
which can be associated with potential or electrostatic point-like interactions, e.g. standard zero-range potential is nothing but the limiting case of electrostatic field barrier.
Other one is given by the BC matrices:
       \begin{equation}\label{eq_bcx2x3}
M_{X_2} =\left(\begin{array}{cc}
               \mu & 0\\
                0 & 1/\mu
            \end{array}
        \right)\,,\qquad
        M_{X_3}=e^{\pi\,i\,\Phi}\left(
        \begin{array}{cc}
          1 & 0 \\
          0 & 1 \\
        \end{array}
      \right)
       \end{equation}
of point-like interactions of magnetic type. Here
\begin{equation}\label{lambda_balian_kurasov}
    X_2 = 2\,\frac{\mu-1}{\mu+1}\,, \quad
    e^{\pi\,i\,\Phi} = \frac{2+i\,X_3}{2-i\,X_3}
\end{equation}
where $\mu = \sqrt{m_{+}/m_{-}}$ is the mass-jump parameter and $\Phi$ is the flux fraction modulo $\pi$.
Magnetic nature of $M_{X_3}$ is obvious because it is interpreted as the localized magnetic flux which breaks the homogeneity of the phase of the wave function $\psi$. Also breaking of the time reversal manifests itself in scattering matrix \cite{qm_zrpsymme_physb2015}.

The natural question arises as to the consideration a particle with internal magnetic moment, e.g. a particle with spin $s=1/2$.
The very straightforward way for derivation of corresponding b.c.
is the conservation of current Eq.~\eqref{j_spinpauli}. Therefore we introduce 4-vector (bispinor) of the the boundary values at the singular point:
\begin{equation}\label{bc_4vector}
  \Phi_{0\pm 0} = \begin{pmatrix}
             \psi_{\uparrow} \\
             \psi'_{\uparrow} \\
             \psi_{\downarrow} \\
             \psi'_{\downarrow} \\
           \end{pmatrix}_{0\pm 0}
\end{equation}
and boundary condition $4\times4$-matrix $M$:
\begin{equation}\label{bc_mgamma4}
  \Phi_{0+0} = M\,\Phi_{0-0}\,.
\end{equation}
Due to the structure of current Eq.~\eqref{j_spinpauli} for the Hamiltonian \eqref{eq_ham_pauli} we have conservation of all its components:
\begin{eqnarray}\label{jxyz_spin}
J_{x}=&\frac{1}{i}\,\left(\Psi^{\dagger}\frac{\partial \Psi}{\partial x}-\frac{\partial \Psi^{\dagger}}{\partial x} \Psi\right)\notag \\
J_{y}=&-\left(\frac{\partial \Psi^{\dagger}}{\partial x} \sigma_{z}\Psi +\Psi^{\dagger}\sigma_{z}\frac{\partial \Psi}{\partial x}\right) \\
J_{z}=&\frac{\partial \Psi^{\dagger}}{\partial x} \sigma_{y}\Psi +\Psi^{\dagger}\sigma_{y}\frac{\partial \Psi}{\partial x}\notag
\end{eqnarray}
Note that here we use expanded form of ``curl`` operator in Eq.~\eqref{j_spinpauli}
with explicit derivatives because we expect the discontinuity in their values. In fact the very this form follows from the Dirac equation in non relativistic limit and the curl-operator appears after collecting the corresponding terms (see \cite{book_ll4_en}). This difference is important in view of the $X_2$ interactions which breaks the homogeneity in dilatation symmetry \cite{funcan_deltasymm_lettmathphys1998} because of the mass jump \cite{qm_zrpsymme_physb2015,qm_deltamassjump_physlett2007}. In general $J_y$ and $J_z$ are non zeroth even if we consider 1-dimensional case, e.g. layered system. The only demand consistent with the hermiticity of the Hamiltonian \eqref{eq_ham_pauli} is the conservation of current components \eqref{jxyz_spin}.

In terms of vector $\Phi$ the components of the probability current are represented as following:
\begin{equation}\label{eq_jxyz}
  J_{i} = \Phi^\dagger\,\Sigma_{i}\Phi\,,\quad i=x,y,z
\end{equation}
where $4 \times 4$ matrices $\Sigma_{i}$ are calculated by comparison of expressions Eq.~\eqref{jxyz_spin} and Eq.~\eqref{eq_jxyz}:
\begin{equation}\label{sigmax_matrices}
  \Sigma_{x} =\frac{1}{i}\begin{pmatrix}
                Sp_2 & 0 \\
                0 & Sp_2
              \end{pmatrix}\,,\quad
              \Sigma_{y} =\begin{pmatrix}
                -\sigma_{x} & 0 \\
                0 & \sigma_{x}
              \end{pmatrix}\,,\quad
\end{equation}
\begin{equation}\label{sigmayz_matrices}
              \Sigma_{z} =\frac{1}{i}\begin{pmatrix}
                0 & \sigma_{x} \\
                -\sigma_{x} & 0
              \end{pmatrix}\, \text{and} \quad
              Sp_2 = \begin{pmatrix}
                0 & 1 \\
                -1 & 0
              \end{pmatrix}\,,
\end{equation}
Thus the conservation of total current gives the conditions for $M$-matrix:
\begin{equation}\label{eq_mmatrix}
    M^{\dagger}\,\Sigma_{i}\,M = \,\Sigma_{i}\,,\,\,i=x,y,z
\end{equation}
%

Besides trivial solution for $M$-matrix consisting of two $M_{X_{2,3}}$-blocks (no spin-flip), simple algebra gives the nontrivial 1-parametric solution of Eqs.~\eqref{eq_mmatrix}:
\begin{equation}\label{mr_matrix}
 M_r =\begin{pmatrix}
                1 & 0 &0 & r \\
                0 & 1&0&0 \\
                0 & r&1&0 \\
                0 & 0&0&1
              \end{pmatrix}\,,\quad r \in \mathbb{R}
\end{equation}
with
\[M_{r_1}\,M_{r_2} = M_{r_1+r_2}\,.\]
and b.c. of the form
\begin{equation}\label{eq_gamma_mr}
\begin{pmatrix}
             \psi_{\uparrow} \\
             \psi'_{\uparrow} \\
             \psi_{\downarrow} \\
             \psi'_{\downarrow} \\
           \end{pmatrix}_{0+0}
           =M_{r}\,\Phi_{0-0} =
           \begin{pmatrix}
           \psi_{\uparrow} +
           r \,\psi'_{\downarrow}\\
           \psi'_{\uparrow}\\
           \psi_{\downarrow}+r\,
              \psi'_{\uparrow}  \\
              \psi'_{\downarrow}\\
              \end{pmatrix}_{0-0}
\end{equation}
which defines the spin-flip variant of $X_4$-extension.
E.g. corresponding scattering matrix for $M_{r}$
is as following:
\begin{equation}\label{eq_scattr}
  \hat{S_{r}} = \frac{1}{k^2\,r^2+4}\,\begin{pmatrix}
  k^2\,r^2 & 4 & -2\,i\,k\,r & 2\,i\,k\,r \\
  4& k^2\,r^2&2\,i\,k\,r&-2\,i\,k\,r\\
  -2\,i\,k\,r& 2\,i\,k\,r& k^2\,r^2 &4 \\
  2\,i\,k\,r &  -2\,i\,k\,r & 4 & k^2\,r^2  \\
                \end{pmatrix}
\end{equation}
\begin{figure}[hbt!]
  \centering
  \includegraphics[scale=0.7]{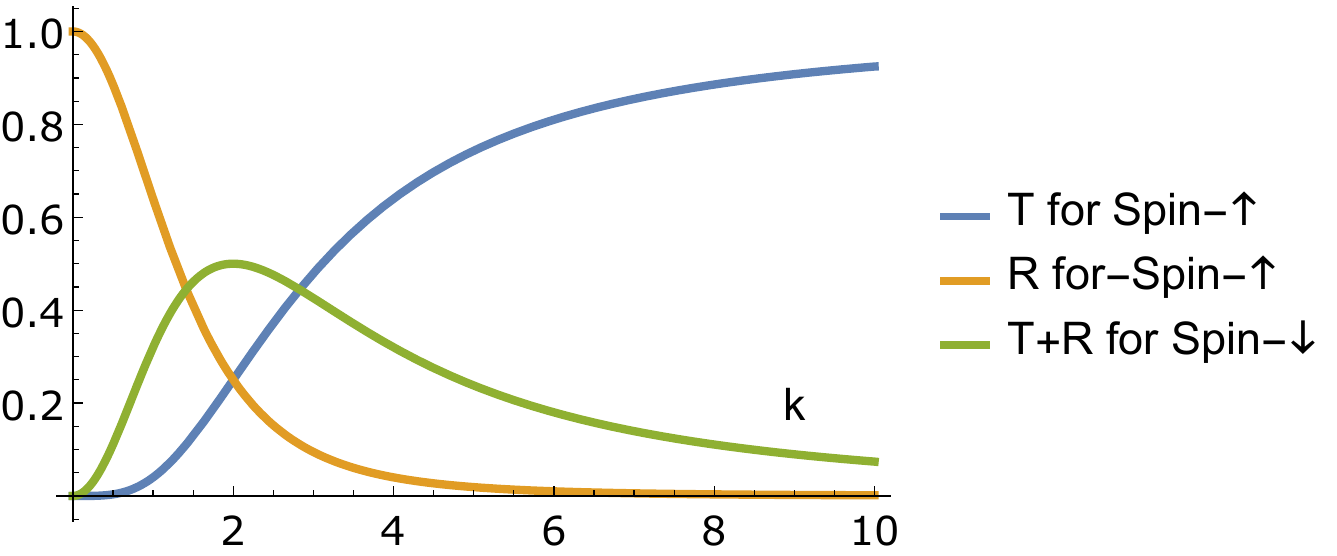}
\caption{Scattering of $\ket{\uparrow}$ - state on $r-X_4$ defect}
\label{fig_scattx4}
\end{figure}

Another solution of Eq.~\eqref{eq_mmatrix} is
\begin{equation}\label{mr1_matrix}
\tilde{M}_{\tilde{r}} =
\begin{pmatrix}
                1 & 0 &0 & 0 \\
                0 & 1&\tilde{r}&0 \\
                0 & 0&1&0 \\
                \tilde{r} & 0&0&1
\end{pmatrix}\,,\quad \tilde{r} \in \mathbb{R}
\end{equation}
with the b.c. of the form:
\begin{equation}
\begin{pmatrix}
             \psi_{\uparrow} \\
             \psi'_{\uparrow} \\
             \psi_{\downarrow} \\
             \psi'_{\downarrow} \\
           \end{pmatrix}_{0+0}
           =
           \tilde{M}_{\tilde{r}}\,\Phi_{0-0} =
 \begin{pmatrix}
  \psi_{\uparrow} \\
 \tilde{r}\, \psi_{\downarrow}+\psi'_{\uparrow} \\
 \psi_{\downarrow} \\
 \tilde{r}\,\psi_{\uparrow}+\psi'_{\downarrow} \\
\end{pmatrix}_{0-0}
\end{equation}
and can be considered as the $\delta$-potential ($X_1$-extension) augmented with the spin-flip mechanism.
From the explicit form of the boundary conditions, e.g.:
\begin{equation}\label{gamma_rashbax2}
\begin{pmatrix}
             \psi_{\uparrow} \\
             \psi'_{\uparrow} \\
             \psi_{\downarrow} \\
             \psi'_{\downarrow} \\
           \end{pmatrix}_{0+0}
           = M_{r}\,M_{X_2}\,\Phi_{0-0} =
           \begin{pmatrix}
           \mu^{-1}\,\psi_{\uparrow} +
           \mu \, r \,\psi'_{\downarrow}\\
           \mu\, \psi'_{\uparrow}\\
           \mu^{-1}\psi_{\downarrow}+\mu \,r\,
              \psi'_{\uparrow}  \\
              \mu\,\psi'_{\downarrow}\\
              \end{pmatrix}_{0-0}
\end{equation}
where $M_{X_2}$ is the block-diagonal matrix of $X_{2}$-extensions. Thus the boundary condition for $s=1/2$ particle with the spin-flip contact interaction can be written in general form:
\begin{equation}\label{bc_mr23_matrix}
\Phi_{0+0} =  \tilde{M}_{\tilde{r}}\,M_{r}\,M_{X_2}\,.
\end{equation}
In contrast to this the $X_3$-extension can not be augmented with the spin-flip mechanism since it trivially decouples from $r$-coupling:
\begin{equation}\label{gamma_rashbax3}
\begin{pmatrix}
             \psi_{\uparrow} \\
             \psi'_{\uparrow} \\
             \psi_{\downarrow} \\
             \psi'_{\downarrow} \\
           \end{pmatrix}_{0+0}
           =M_{r}\,M_{X_3}\,\Phi_{0-0} = e^{i\,\pi\,\Phi}\,
           \begin{pmatrix}
           \psi_{\uparrow} +
           r \,\psi'_{\downarrow}\\
           \psi'_{\uparrow}\\
           \psi_{\downarrow}+r\,
              \psi'_{\uparrow}  \\
              \psi'_{\downarrow}\\
              \end{pmatrix}_{0-0}
\end{equation}
In accordance with the spin-momentum nature of the $r$-couplings the physical reason of such factorization is that $X_3$ contact interaction does not include spatial inhomogeneity in electric field potential $\varphi$. This is quite consistent with the difference between $X_2$ and $X_3$ from the point of view of breaking the gauge symmetry \cite{qm_zrpsymme_physb2015,qm_deltamassjump_us_arxiv2018}.

\begin{figure}[hbt!]
\centering
\includegraphics[scale=0.7]{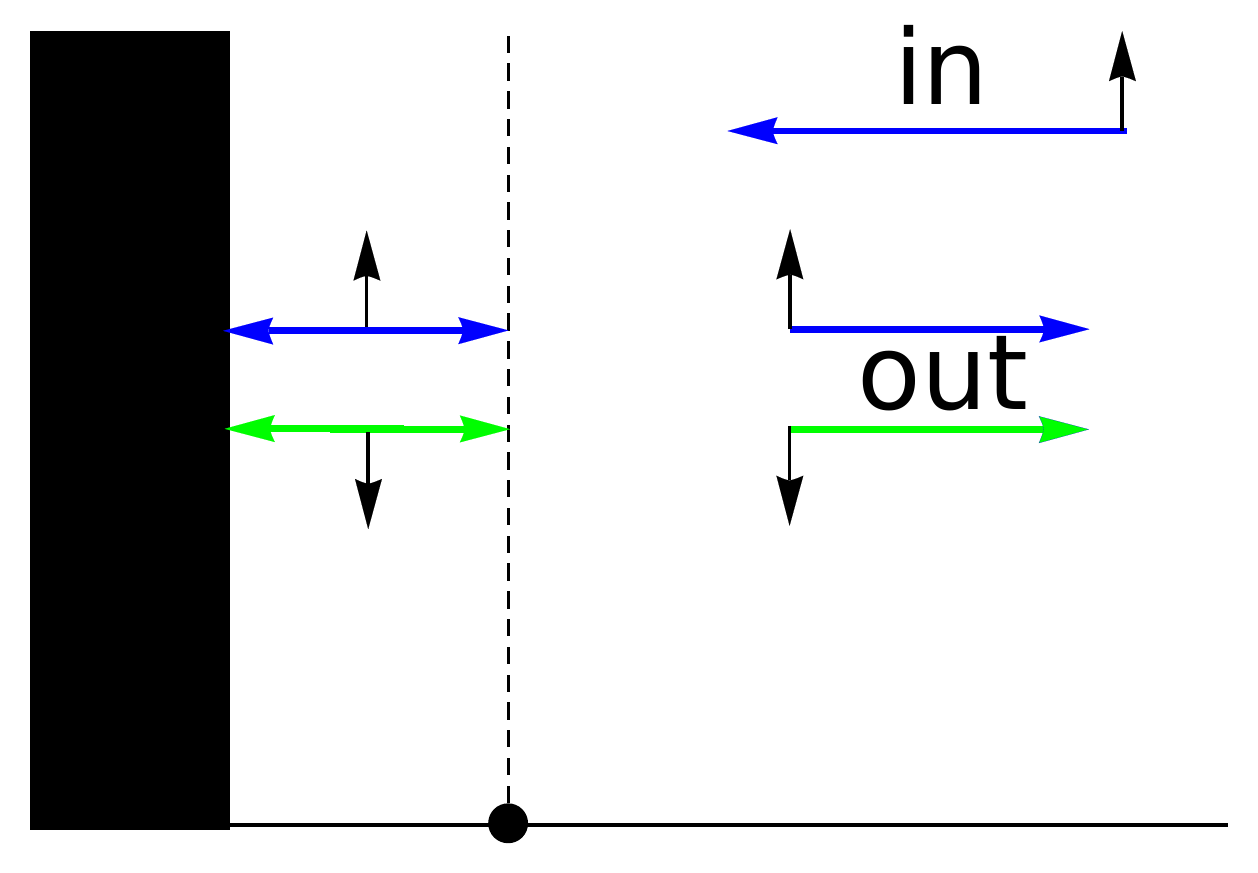}
\caption{Resonator}\label{fig_res}
\end{figure}

\begin{figure}[hbt!]
\centering
        \includegraphics[width=0.45\textwidth]{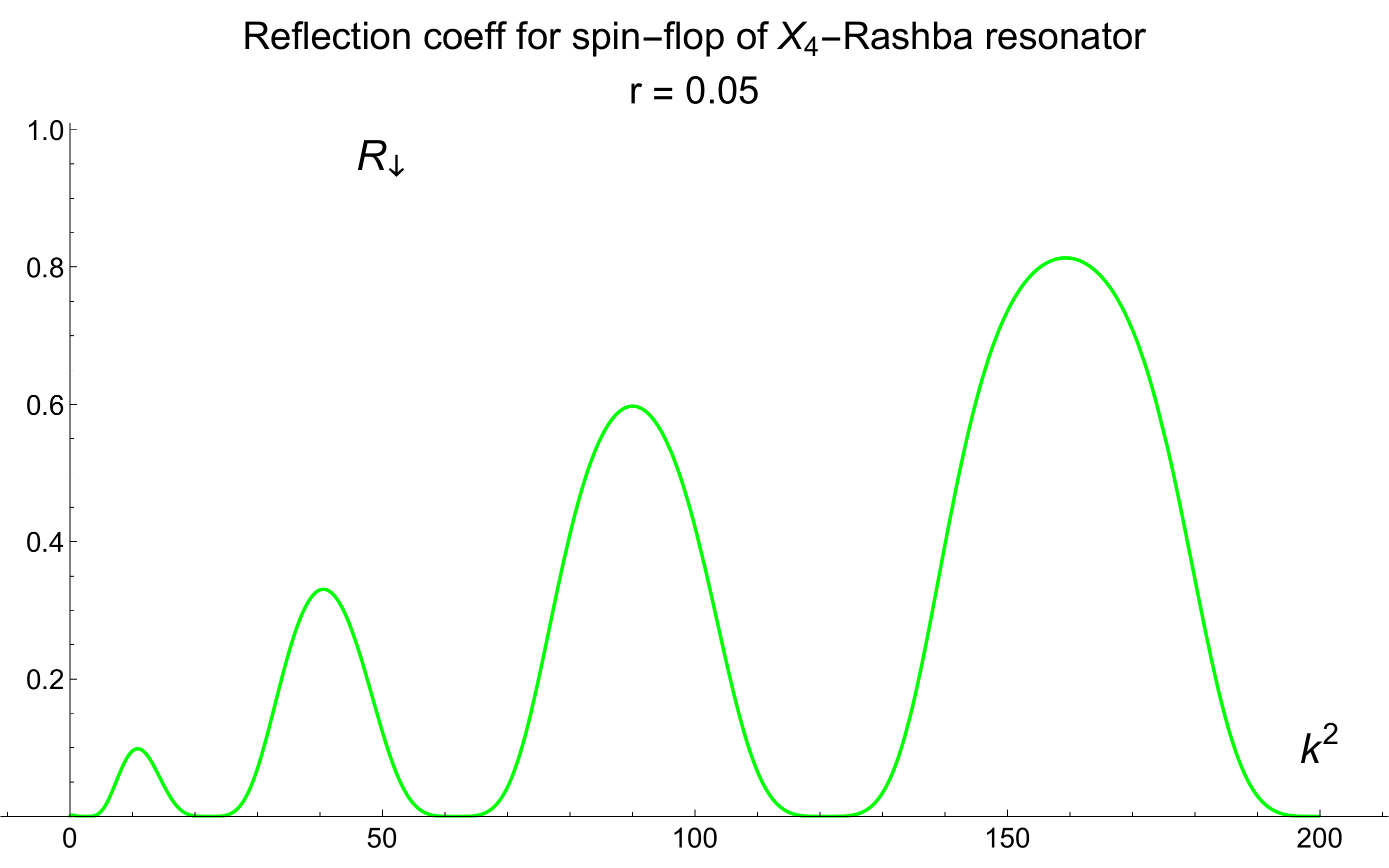}
\,\,
        \includegraphics[width=0.45\textwidth]{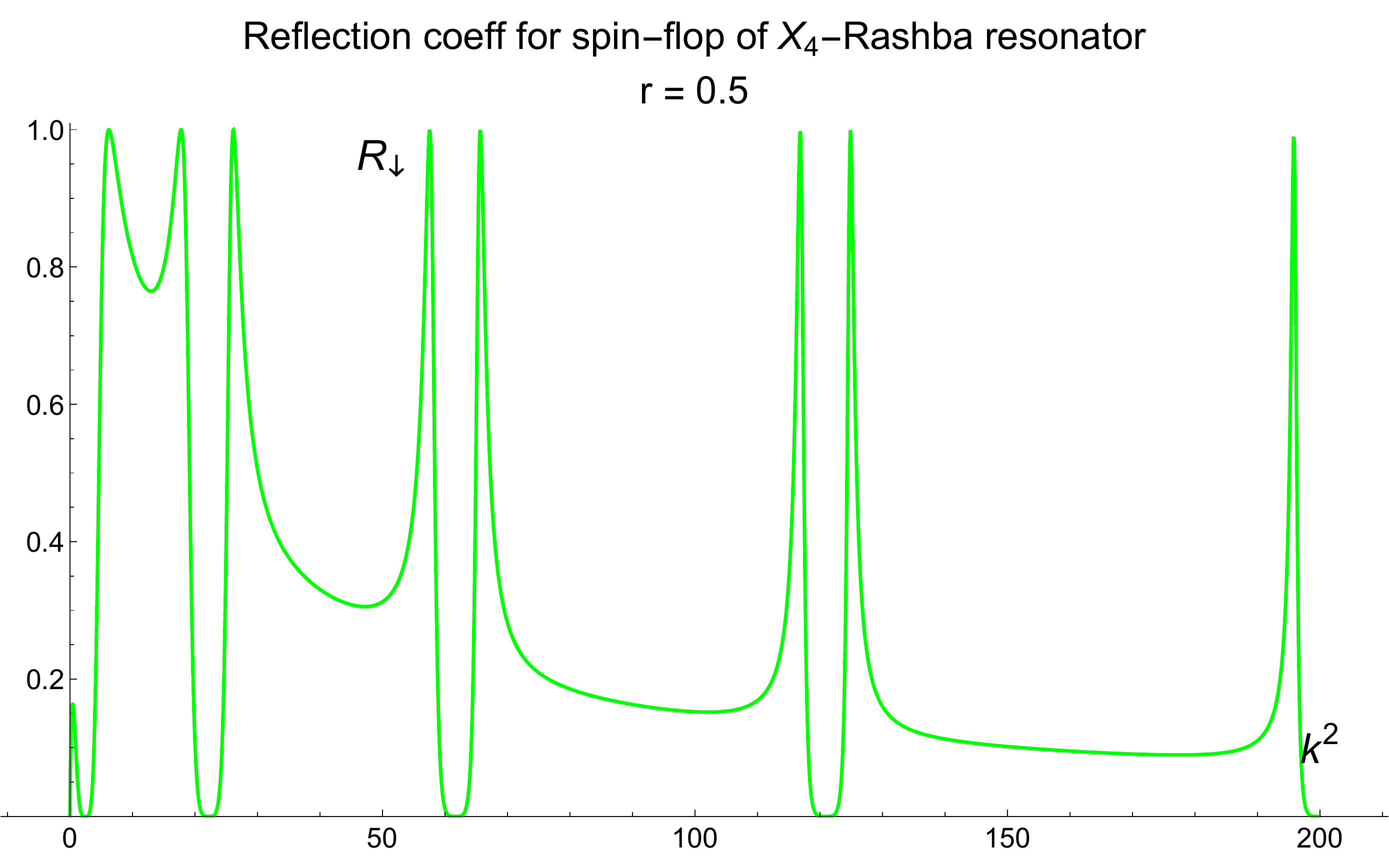}
       \caption{Intensity of reflected spin-$\downarrow$ state for $r-X_4$ resonator (see Fig.~\ref{fig_res}) at different values of $r$.}\label{fig_resr}
\end{figure}

\begin{figure}
\centering
\includegraphics[scale=0.7]{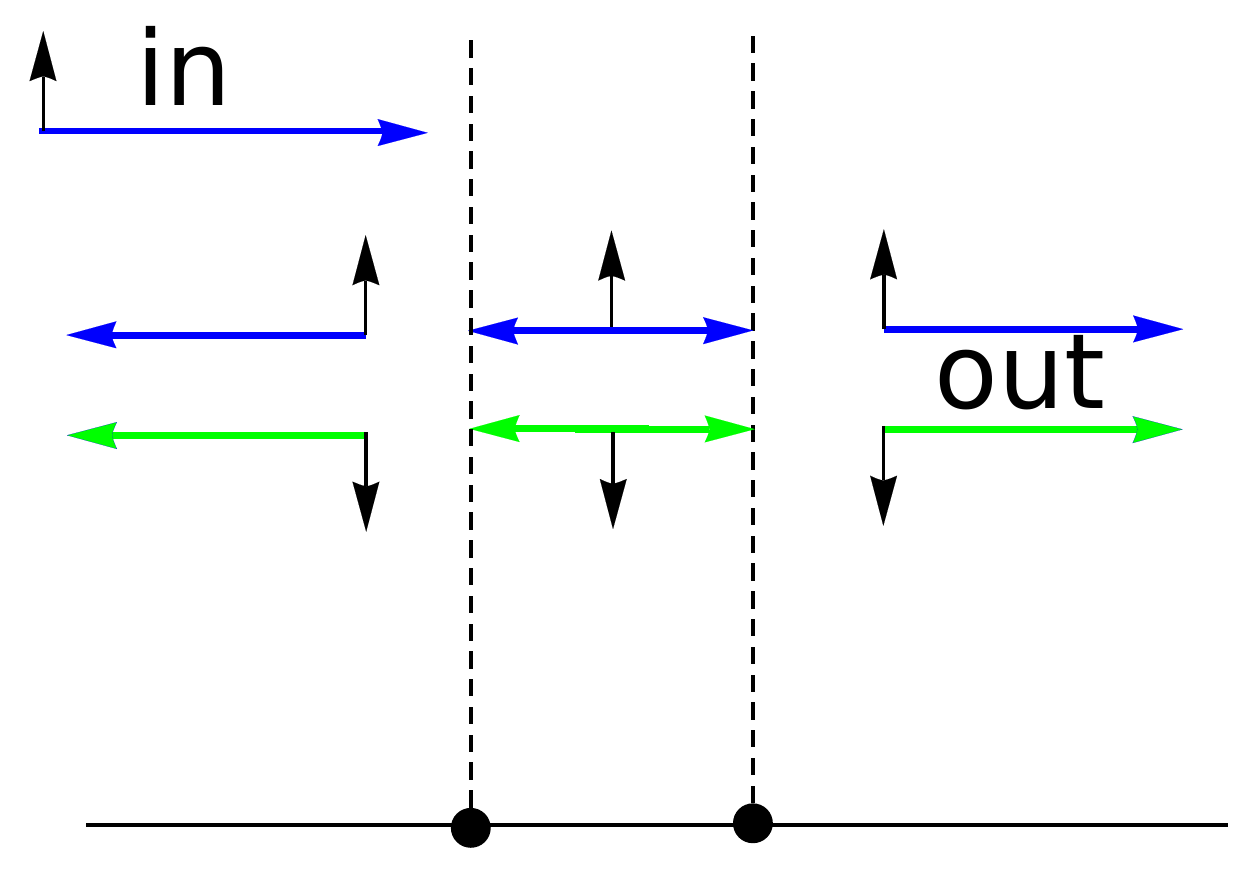}
\caption{Filter}\label{fig_filter}
\end{figure}

\begin{figure}[hbt!]
\centering
        \includegraphics[width=0.45\textwidth]{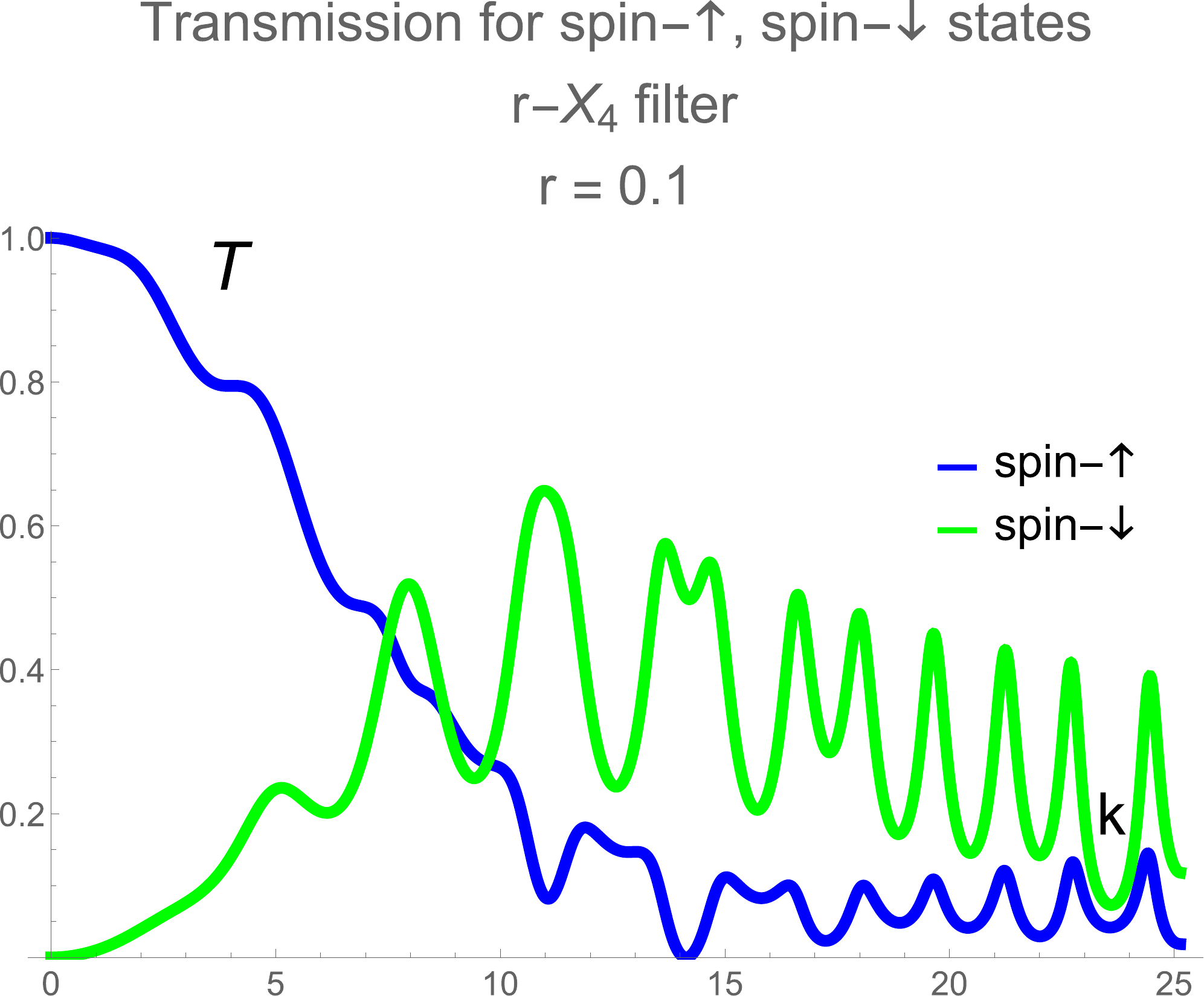}
\,\,
        \includegraphics[width=0.45\textwidth]{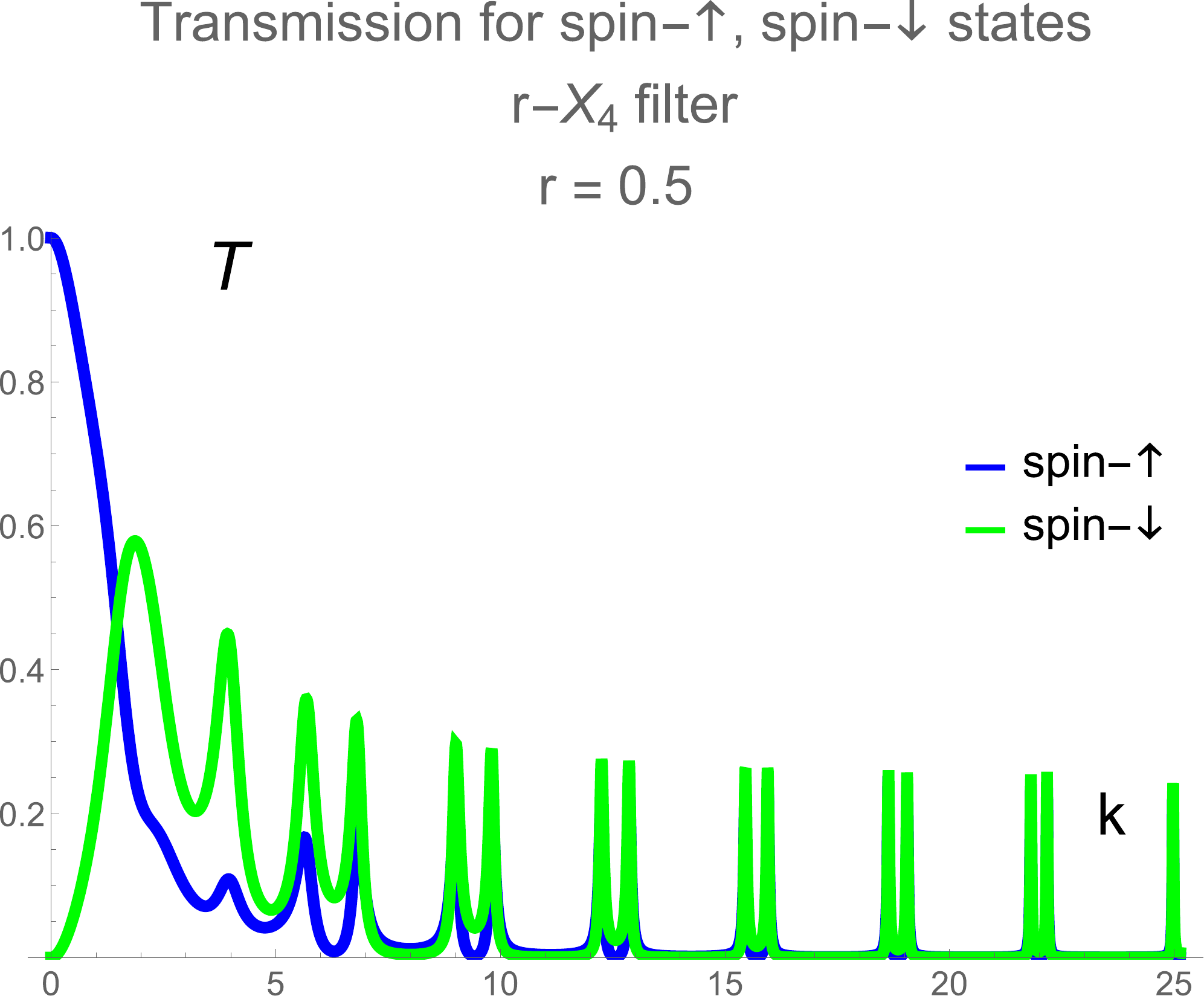}
       \caption{Transmission $r-X_4$ filter intensity for different values of $r$.}\label{fig_filterr}
\end{figure}

On this basis the standard test systems and their transport characteristics can be calculated straightforwardly in order to demonstrate spin-filtering properties. We give here just two examples here: the resonator (see Fig.~\ref{fig_res},\ref{fig_resr}), and the filter (see Fig.~\ref{fig_filter},\ref{fig_filterr}).
The intensity of spin-flip process, generating the spin-$\downarrow$ state from incident spin-$\uparrow$ state is shown in Fig~\ref{fig_resr}. These results demonstrate that spin-flip mechanism even at small values of $r$-coupling can reach high probabilities with increasing the energy of incident particle. Of course this directly follows from the boundary conditions \eqref{mr_matrix} and \eqref{mr1_matrix} since the effects depend on both $r$ and the momentum.
\begin{figure}[hbt!]
\centering
        \includegraphics[width=0.45\textwidth]{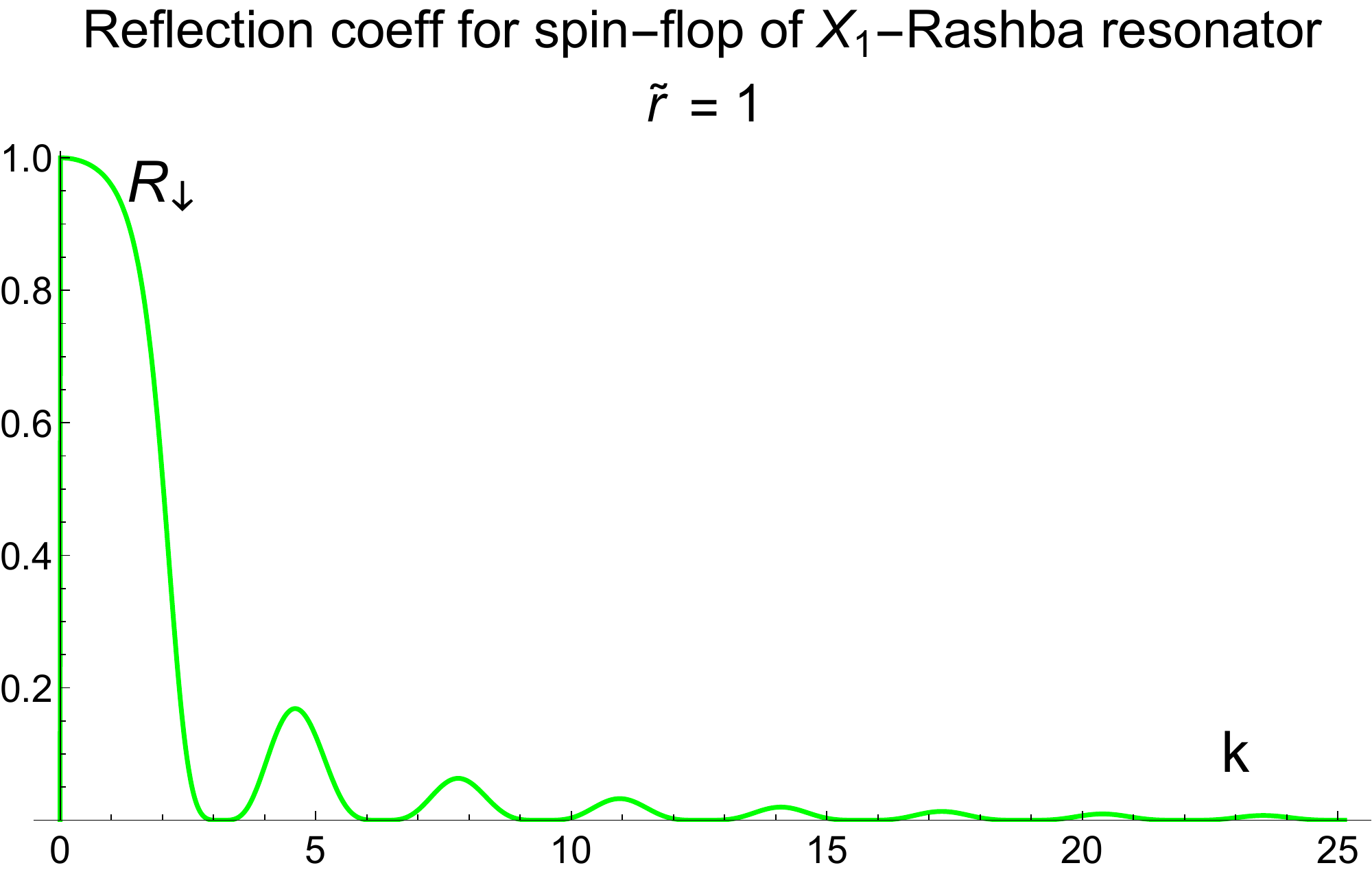}
\,\,
        \includegraphics[width=0.45\textwidth]{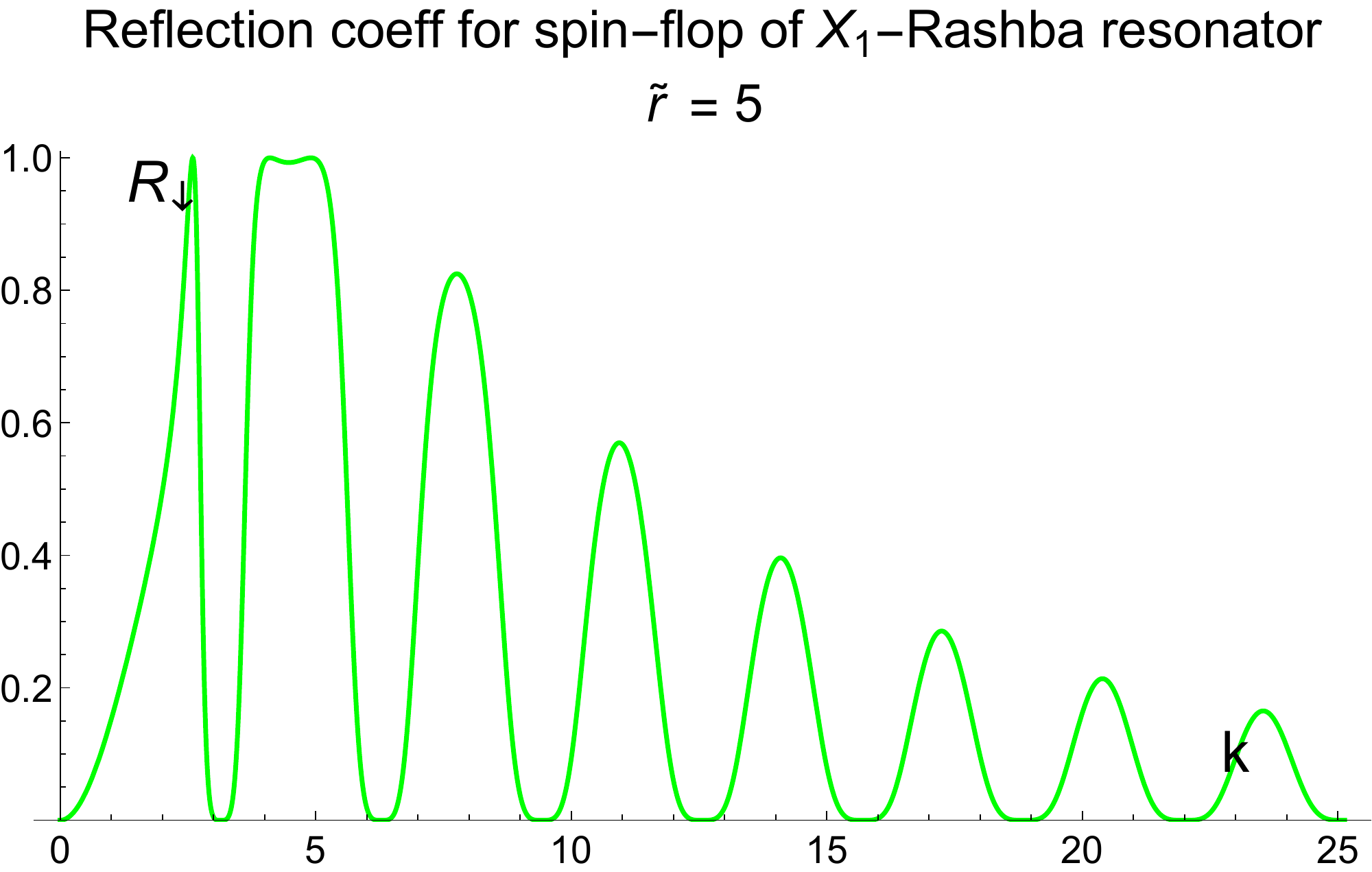}
       \caption{Intensity of reflected spin-$\downarrow$ state  for $\tilde{r}-X_1$ resonator (see Fig.~\ref{fig_res}) at different values of $\tilde{r}$.}\label{fig_resrx1}
\end{figure}
Comparison of $\tilde{r}-X_1$ and $r-X_4$ cases shows that the last one is more effective as spin-flipping mechanism.

The zone structure for $r-X_4$ periodic comb can be also calculated in standard way. It strongly depends on $r$. The lowest states belong to two parabolic zones with different effective mass at $r<1$:
\begin{equation}\label{eq_displaw_mass}
  E_{\pm}(k) = \frac{\hbar^2\,k^2}{2\,m_{\pm}}\,,\quad
  m_{\pm} = 1\pm r
\end{equation}
%
%
At $r=1$ one branch of excitations becomes massless $E(k) = 2\sqrt{3}\,k$. Of course this is the remnant of what happens in standard $X_4$-structure \cite{math_deltalbergestexactsolv}. More intriguing problem here is the inclusion of the correlation effects due to spin statistics and investigation of phases with magnetic (dis)order in dependence on the intensity of point-like interactions. This way of research may be useful for modeling 1-dimensional magnetic systems \cite{qm_1Dmagnetism_lnph2004}.

\section{Physical origin of the spin-flip boundary conditions}\label{sec_spinrashba}
The spin-flip point interactions introduced above make the spin operator no longer the integral of motion. There are two obvious physical origins for it a) an external magnetic field with $x,y$-components and b) spin-momentum coupling (Rashba coupling). The explicit $k$-dependence of the amplitudes of the spin-flip processes indicates that these interactions are due to spin-momentum coupling. Thus the physical interpretation of interactions represented by the b.c. matrices $M_r, \tilde{M}_{\tilde{r}}$ can be given
in terms of
the Rashba Hamiltonian
\cite{qm_rashbaorigin_physolid1960,qm_rashbabychkov_zhetp1984} (see also \cite{qm_rashbaham_nat2015} and reference therein).
Indeed, the Pauli Hamiltonian Eq.~\eqref{eq_ham_pauli} as well as the current density Eq.~\eqref{j_spinpauli} can be derived as the non relativistic limit for the Dirac Hamiltonian
\begin{equation}\label{eq_hamdirac}
  \hat{H}_{D} =\boldsymbol{\alpha}\cdot \left(\hat{\mathbf{p}}-\mathbf{A}\right)+\beta\,m + \varphi
\end{equation}
where $\boldsymbol{\alpha} = \alpha_{i},\,i=1,2,3$ and $\beta$ are the Dirac matrices
\begin{equation}
\boldsymbol{\alpha} = \begin{pmatrix}
0& \boldsymbol{\sigma}\\
\boldsymbol{\sigma}&0
\end{pmatrix}\,,\quad
\beta = \begin{pmatrix}
I& 0\\
0&-I
\end{pmatrix}
\end{equation}
with $I$ being $2\times2$ unit matrix.
They act in the space of bispinors $\Psi$:
\begin{equation}\label{eq_bispinor}
 \Psi_{D} = \begin{pmatrix}
             \xi\\
             \eta
           \end{pmatrix}
\end{equation}
where spinors $\xi$ and $\eta$ represent particle and hole with respect to the Dirac vacuum states respectively \cite{book_ll4_en}. The probability density is:
\begin{equation}\label{eq_jdirac}
    \mathbf{J}_{D}=\Psi_{D}^\dag \,\boldsymbol{\alpha}\, \Psi_{D}
\end{equation}
and in non relativistic limit transforms into
\begin{equation}
    \mathbf{J}=\xi^* \,\boldsymbol{\sigma}\, \eta +\eta^* \,\boldsymbol{\sigma}\, \xi
\end{equation}
with
\begin{equation}
    \eta = \frac{1}{2\,m}\hat{\mathbf{v}}\,\xi
\end{equation}
Here $\hat{\mathbf{v}}$ is the velocity operator.
In the absence of external electromagnetic field this is equivalent to the following reduction of the bispinor in 1-dimensional case
\begin{equation}\label{eq_bispinor2spinor}
\Psi_{D}\to \begin{pmatrix}
             \xi\\
             \xi'
           \end{pmatrix}
\end{equation}
so that the boundary element 4-vector \eqref{bc_4vector} appears. Also we refer to the paper \cite{qm_deltamassjumpdirac_jmph2015} where mass jump matching conditions were derived for the Dirac Hamiltonian in a graphen-like material where the speed of light interchanged with the Fermi velocity $v_{F}$.

The expansion of next order generates the spin dependent operator in the Hamiltonian:
\begin{equation}\label{eq_spinmomcoupl}
  \hat{H}_{SP} = \lambda\,\boldsymbol{\sigma}\cdot \left(\nabla \varphi \times \hat{\mathbf{p}}\right)
\end{equation}
It couples the spin with the momentum due to inhomogeneous background of the electric potential $\varphi$. In the limiting case of point-like interaction on the axis when $\nabla \varphi \to 0$ on both sides of the singular point this term drops out and should be interchanged with the boundary condition for the corresponding boundary vector \eqref{bc_4vector} of the Pauli Hamiltonian \eqref{eq_ham_pauli}. The conservation of the corresponding probability density current Eq.~\eqref{j_spinpauli} provides self-adjointess of the boundary conditions for \eqref{eq_ham_pauli} in the presence of point-like singularity.

As a result, all
extensions $X_{i}\,,\,\,i=1,2,4$ which are singular limiting cases of the spatial distribution of the external electric field potential $\varphi$ can be augmented with the spin-flip mechanism.
Thus Eq.~\eqref{bc_mr23_matrix} defines the one-dimensional
analog of the Hamiltonian with the point-like Rashba spin-momentum interaction  \cite{qm_rashbaorigin_physolid1960}.

\section*{Conclusion}
The main result of the paper is that those extensions of the Schr\"{o}dinger operator which are physically constracted on the basis of the inhomogeneous distribution of the electric field potential $\varphi(x)$ can be augmented with the spin-flip mechanism.
Note that in Eq.~\eqref{gamma_rashbax2} both $r$-coupling and $\mu$-parameter determine
the spin-flip mechanism. This is in coherence with the results of \cite{qm_deltamassjump_us_arxiv2018} where $X_2$ and $X_4$ extensions were treated on the common basis of the spatial dependent effective mass. In its turn it is caused by the electrostatic field of the crystalline background. So it is not surprise that these extensions can be combined through
spin-momentum coupling in the Rashba Hamiltonian thus forming the ``internal``
magnetic field. In contrast to this pure ``magnetic`` $X_3$-extension which is due to the external magnetic field does not couple with the Rashba point-like interactions.

Thus we can state that one-dimensional analog of the Rashba Hamiltonian is obtained.
It is interesting to check this result independently using the Kurasov's distribution theory technique \cite{funcan_deltadistr_kurasov_jmathan1996} modified correspondingly for spin $1/2$ case. This will be done in future work. We

The authors thank Prof.~Vadim Adamyan for clarifying comments and discussions. This work was completed due to individual (V.K.) Fulbright Research Grant (IIE ID: PS00245791) and with support by MES of Ukraine, grants 0115U003208 and 018U000202. V.K. is also grateful to Mr.~Konstantin Yun for financial support of the research.

The author declare that there is no conflict of
interest.

\newpage
\section*{References}
\providecommand{\newblock}{}

\end{document}